\documentclass[sigconf,screen]{acmart} 

\usepackage{algorithm}
\usepackage[noend]{algpseudocode}
\usepackage{algorithmicx}
\usepackage{graphicx}
\usepackage{textcomp}
\usepackage{xcolor}

\usepackage{makecell}
\usepackage{array}
\usepackage{booktabs}

\usepackage{xcolor, soul}
\usepackage{xspace}
\usepackage{multirow}
\usepackage{tabularx}
\usepackage{tikz}
\usepackage[bottom]{footmisc}

\usepackage{mathtools}

\usepackage{subcaption}
\usepackage{booktabs}

\usepackage{pgfplots}
\pgfplotsset{compat=1.17} 

\usepackage{stfloats}

\usepackage{enumitem}

\usepackage{mwe}

\definecolor{lavenderr}{rgb}{0.71, 0.49, 0.86}

\usepackage{graphicx}

\usepackage{tikzducks}

\usepackage[en-GB]{datetime2} 
\DTMsetup{
    showzone=false,   
    showseconds=false 
}

\definecolor{darkspringgreen}{rgb}{0.09, 0.45, 0.27}
\definecolor{denim}{rgb}{0.08, 0.38, 0.74}
\definecolor{darkolivegreen}{rgb}{0.33, 0.42, 0.18}
\definecolor{tangerine}{rgb}{0.95, 0.52, 0.0}
\definecolor{mahogany}{rgb}{0.75, 0.25, 0.0}
\definecolor{coolblack}{rgb}{0.0, 0.22, 0.44}
\definecolor{darkpink}{rgb}{0.91, 0.35, 0.6}
\definecolor{darkblue}{rgb}{0.0, 0.0, 0.67}
\definecolor{melon}{rgb}{0.97, 0.69, 0.67}

\definecolor{seagreen}{rgb}{0.18, 0.55, 0.34}
\definecolor{pred}{rgb}{0.7843, 0.0039, 0.3137} 

\newcommand{\kk}[1]{{\color{black}{#1}}} 
\newcommand{\kktwo}[1]{{\color{black}{#1}}} 
\newcommand{\kkthree}[1]{{\color{black}{#1}}} 

\definecolor{darkpink}{rgb}{0.88, 0.28, 0.54}
\definecolor{forestgreen}{rgb}{0.0, 0.27, 0.13}
\definecolor{amber}{rgb}{1.0, 0.49, 0.0}

\newcommand{\inum}[1]{(\textit{#1})\xspace}

\newcommand{\head}[1]{{\vspace{2pt}\noindent\textbf{#1.}\xspace}} 

\newcolumntype{Y}{>{\centering\arraybackslash}X}
\usepackage{tikz}

\newcommand{\squishlist}{
 \begin{list}{$\circ$}
  { \setlength{\itemsep}{0pt}
     \setlength{\parsep}{0pt}
     \setlength{\topsep}{3pt}
     \setlength{\partopsep}{0pt}
     \setlength{\leftmargin}{1em}
     \setlength{\labelwidth}{1em}
     \setlength{\labelsep}{0.5em} } }

\newcommand{\squishend}{
  \end{list}  }

\makeatletter
\g@addto@macro{\normalsize}{%
  \setlength{\abovedisplayskip}{4pt plus 0.5pt minus 1pt}
  \setlength{\belowdisplayskip}{4pt plus 0.5pt minus 1pt}
  \setlength{\abovedisplayshortskip}{0pt}
  \setlength{\belowdisplayshortskip}{0pt}
  \setlength{\intextsep}{3pt plus 1pt minus 1pt}
  \setlength{\textfloatsep}{7pt plus 1pt minus 1pt}
  \setlength{\skip\footins}{4pt plus 1pt minus 1pt}}
\makeatother

\usepackage{blindtext,graphicx}
\usepackage[absolute]{textpos}
\usepackage{datetime}

\definecolor{seagreen}{rgb}{0.18, 0.55, 0.34}
\definecolor{ballblue}{rgb}{0.13, 0.67, 0.8}

\definecolor{darkgreen}{rgb}{0.0, 0.44, 0.34}

\sloppypar

\definecolor{dollarbill}{rgb}{0.52, 0.73, 0.4}

\usetikzlibrary{patterns}
\usepackage[precision=2, unit=mm]{lengthconvert}

\usepackage[colorinlistoftodos,prependcaption,textsize=scriptsize]{todonotes}
\usepackage{marginnote} 
\definecolor{cyan(process)}{rgb}{0.0, 0.62, 0.82}

\definecolor{cadmiumgreen}{rgb}{0.0, 0.50, 0.29}

\newcommand\revref[1]{\hyperref[rev:#1]{#1}}

\definecolor{raspberry}{rgb}{0.89, 0.04, 0.36}

\definecolor{awesome}{rgb}{1.0, 0.13, 0.32}
\definecolor{cardinal}{rgb}{0.77, 0.12, 0.23}
\definecolor{cadet}{rgb}{0.33, 0.41, 0.47}
\definecolor{celadon}{rgb}{0.67, 0.88, 0.69}
\definecolor{persianblue}{rgb}{0.11, 0.22, 0.73}
\definecolor{ultramarine}{rgb}{0.07, 0.04, 0.56}
\definecolor{warmblack}{rgb}{0.0, 0.3, 0.3}

\definecolor{darkpastelgreen}{rgb}{0.01, 0.75, 0.24}
\newcommand\nmg[1]{{\color{black}{#1}}}
\newcommand\nmgg[1]{{\color{black}{#1}}}

\definecolor{terracotta}{rgb}{0.89, 0.45, 0.36}

\definecolor{forestgreen(web)}{rgb}{0.13, 0.55, 0.13}

\renewcommand{\figurename}{Fig.}

\definecolor{cardinal}{rgb}{0.77, 0.12, 0.23}
\definecolor{deeppink}{rgb}{1.0, 0.08, 0.58}
\definecolor{brightpink}{rgb}{1.0, 0.0, 0.5}
\definecolor{electricviolet}{rgb}{0.56, 0.0, 1.0}
\definecolor{brandeisblue}{rgb}{0.0, 0.44, 1.0}
\definecolor{carminered}{rgb}{1.0, 0.0, 0.22}

\definecolor{acolor}{rgb}{0.0, 0.5, 1.0}
\definecolor{bcolor}{rgb}{0.54, 0.17, 0.89}
\definecolor{ccolor}{rgb}{0.4, 0.69, 0.2}
\definecolor{dcolor}{rgb}{0.92, 0.41, 0.12}
\definecolor{ecolor}{rgb}{0.6, 0.0, 0.156}
\definecolor{fcolor}{rgb}{0.106, 0.620, 0.467}

\definecolor{dogwoodrose}{rgb}{0.84, 0.09, 0.41}

\newcommand\omcr[1]{{\color{black}{#1}}}

\newcommand{\citesbs}{bentley_accurate_2008,margulies_genome_2005,shendure_accurate_2005,harris_single-molecule_2008,turcatti_new_2008,wu_termination_2007,fuller_rapid_2007,mckernan_reagents_2008,fuller_method_2011}
\newcommand{\citesmrt}{eid_real-time_2009}
\newcommand{\citenanopore}{menestrina_ionic_1986,cherf_automated_2012,manrao_reading_2012,laszlo_decoding_2014,deamer_three_2016,kasianowicz_characterization_1996,meller_rapid_2000,stoddart_single-nucleotide_2009,laszlo_detection_2013,schreiber_error_2013,butler_single-molecule_2008,derrington_nanopore_2010,song_structure_1996,walker_pore-forming_1994,wescoe_nanopores_2014,lieberman_processive_2010,bezrukov_dynamics_1996,akeson_microsecond_1999,stoddart_nucleobase_2010,ashkenasy_recognizing_2005,stoddart_multiple_2010,bezrukov_current_1993,zhang_single-molecule_2024}

\newcommand{\citepersonalized}{alkan_personalized_2009,lightbody_review_2019,morganti_next_2019,branco_bioinformatics_2021,quazi_artificial_2022,aronson_building_2015,f_lochel_comparative_2020,papadopoulou_application_2023,tafazoli_applying_2021,gambardella_personalized_2020,leary_development_2010,hamburg_margaret_a_path_2010,van_der_lee_technologies_2020,moon_precision_2022,mohan_profiling_2020,chung_rapid_2020,bielinski_preemptive_2014,ho_enabling_2020,hussen_emerging_2022,russell_pharmacogenomics_2021,verma_nanopore_2024,clark2019diagnosis,farnaes2018rapid,sweeney2021rapid,flores2013p4,ginsburg2009genomic,chin2011cancer,Ashley2016}

\newcommand{\citeoutbreakrapid}{dunn_squigglefilter_2021,bertelli_rapid_2013,arias_rapid_2016,comin_investigation_2020,Quick2016}

\newcommand{\citeoutbreak}{\citeoutbreakrapid,robinson_genomics_2013,fournier_clinical_2014,koser_routine_2012,eloit_diagnosis_2014,gardy_jennifer_l_whole-genome_2011,taylor_angela_j_characterization_2015,quainoo_scott_whole-genome_2017,goldberg_brittany_making_2015,besser_interpretation_2019,li_application_2021,deng_integrated_2021,kwong_whole_2015,deurenberg_application_2017,tang_infection_2017,croucher_application_2015,
bloom2021massively,yelagandula2021multiplexed,le2013selected,nikolayevskyy2016whole,qiu2015whole,gilchrist2015whole}

\newcommand{\citeagriculture}{the_arabidopsis_genome_initiative_analysis_2000,zhu_applications_2020,choi_nanopore_2020,stevens_sequence_2016,campos_high_2021,gao_genome_2021,van_dijk_machine_2021,sun_twenty_2022,kim_application_2020,thudi_genomic_2021,michael_building_2020,shen_omics-based_2022,shahroodi_demeter_2022,
prasad2021soil,Mascher2024,Schreiber2024}

\newcommand{\citeevolution}{kanehisa_toward_2019,qing_whole_2022,wittkopp_cis-regulatory_2012,romero_comparative_2012,wang_population_2020,hill_molecular_2021,vaishnav_evolution_2022,zhang_haplotype-resolved_2021,fay_evaluating_2008,kanzi_next_2020,wray_evolution_2003,wu_one_2021,signor_evolution_2018,whitehead_variation_2006,coolon_tempo_2014,
ellegren2014genome,Prado-Martinez2013,Prohaska2019human}

\newcommand{\citecancer}{lawrence_mutational_2013,vogelstein_cancer_2013,ramskold_full-length_2012,baslan_unravelling_2017,shapiro_single-cell_2013,sakamoto_new_2020,jia_high-throughput_2022,lawson_tumour_2018,liu_mrna-based_2023,van_de_sande_applications_2023,chakravarty_clinical_2021,cortes-ciriano_computational_2022,deveson_evaluating_2021,xiao_toward_2021,bolton_cancer_2020,szustakowski_advancing_2021,navin_future_2011,hong_rna_2020,lei_applications_2021,han_single-cell_2022,federici_variants_2020,zhang_singlecell_2021,ren_understanding_2018,tian_cicero_2020,malone_molecular_2020,tang_single-cell_2019,ellsworth_single-cell_2017,zhong_application_2021,stadler_therapeutic_2021,tan_targeted_2022,degasperi_substitution_2022,xu_single-cell_2022,horak_comprehensive_2021,zhang_single-cell_2016,bruno_next_2020,de_luca_fgfr_2020,waarts_targeting_2022,lim_advancing_2020,colomer_when_2020,saadatpour_single-cell_2015,dizman_sequencing_2020,buzdin_rna_2020,xiao_tumor_2021,nandwani_lncrnas_2021,marchetti_error-corrected_2023,chen_next-generation_2021,navin_first_2015}

\newcommand{\citealgoptimization}{zhang2000greedy,slater2005automated,li2018minimap2,myers1999fast,marco2021fast,marcosola2023optimal,grootkoerkamp2024apa2,xin2013accelerating,xin2015shifted,sadasivan2024genomic,tseng2025ultrafast,walia2025ultrafast,kim_fastremap_2022,Ashyralyyev2026gencore,ashyralyyev2026lcpan,alicioglu2024pairwise,manber1993suffix}
\newcommand{\citegraphalgoptimization}{rautiainen2020graphaligner,kim2019hisat2,gao2020abpoa,jain2019pasgal,siren2021pangenomics,Rautiainen2019,Chandra2023,Ivanov2022,Ma2023,Darby2020vargas,Hwang2025MEMO,Romain2023svjedi,Li2020minigraph}

\newcommand{\citehwoptimization}{doblas2025smx,mutlu2023accelerating,alser2022molecules,lou2020helix,lou2018brawl,shahroodi2023swordfish,markus2020benchmarking,subramaniyan2021accelerated,huangfu2018radar,khatamifard2021genvom,gupta2019rapid,li2021pim,angizi2019aligns,zokaee2018aligner,turakhia2018darwin,fujiki2018genax,madhavan2014race,cheng2018bitmapper2,houtgast2018hardware,houtgast2017efficient,zeni2020logan,ahmed2019gasal2,nishimura2017accelerating,de2016cudalign,liu2015gswabe,liu2013cudasw++,liu2009cudasw++,liu2010cudasw++,wilton2015arioc,goyal2017ultra,chen2016spark,chen2014accelerating,chen2021high,fujiki2020seedex,banerjee2018asap,fei2018fpgasw,waidyasooriya2015hardware,chen2015novel,rucci2018swifold,haghi2021fpga,li2021pipebsw,ham2020genesis,ham2021accelerating,wu2019fpga,cali2020genasm,Zhang_2023_alignerD,soysal2025mars,kim2018grim,kaplan2020bioseal,mao2022genpip,dphls2026,wang20202,Walia2024talco,sadasivan2024genomic,Turakhia2025toward,Turakhia2019darwinwga,simon2026processing,eudine2026genpairx,Lindegger2023scrooge}

\newcommand{\citegraphhwoptimization}{cali2022segram,Zhang2024Harp,Zeng2024asgdp,Shen2024128parallel,Li2024,Mandal2020,Varma2013,Awan2021,Feng2021,Zhang2025,kim2025nmp,Huang2023meg2,Angizi2020Panda,Qiu2017,Zhou2021,Sarkar2021,Varma2017,Varma2016,Goswami2018,Galanos2021,Angizi2020,Sinha2022,Meng2014,Hu2016,Chen2023,Natarajan2018,Ren2018}

\newcommand{\citetranscriptomics}{wang2009rna,lowe2017transcriptomics,angerer2017single,Stark2019,chen2023hitchhikers,weirather2017comprehensive,Sibbesen2023,haas_forensic_2021,liu_desalt_2019,lachmann2018massive,clough2023ncbigeo,bray2016near}

\renewcommand\footnotetextcopyrightpermission[1]{}
\setcopyright{none}

\begin{document}

\title{Architecture for Health Initiative (Arch4Health):\\
Computational Challenges in Health-Related Applications\\ and the Role of Computer Architecture in Addressing Them}



\def\iscacameraready{} 
\newcommand{\hpcapubid}{0000--0000/00\$00.00}



\setcounter{page}{1}

\begin{abstract}
Recent biotechnological advances enable high-throughput, low-cost, and accurate biological data generation. This wealth of data enables unique opportunities for advancing healthcare. Despite these opportunities, efficiently analyzing large-scale biological data poses significant challenges for conventional computing systems. These systems often cannot keep up with the high-throughput rate at which data is generated, and they face additional constraints related to energy efficiency, scalability, privacy, and security. Therefore, to facilitate the wide adoption of recent advances in healthcare, there is a need to optimize the computing systems to enable high-performance, energy-efficient, low-cost, private, and secure analysis of biological data. 

We introduce the \textbf{Architecture for Health (Arch4Health)} initiative, which aims to \inum{i}~identify \kk{and analyze} key computational challenges in \kk{current and future} \kk{health- and life science-related} applications and \inum{ii}~explore how computer architects \kk{and computing system designers} can advance healthcare by addressing these challenges. In this short paper, we first present the motivations behind the Arch4Health initiative and, second, elaborate on its vision and goals, related topics, Arch4Health workshops, and future outlooks.
 
\end{abstract}
\pagestyle{plain}   

\author{Nika Mansouri Ghiasi\nmg{*}}
\affiliation{%
\authornote{Both authors contributed equally to this paper.}
  \institution{ETH Zürich}
  \country{Switzerland}
}

\author{Konstantina Koliogeorgi\nmg{*}}
\affiliation{%
\authornotemark[1]
  \institution{ETH Zürich}
  \country{Switzerland}
}

\author{Onur Mutlu}
\affiliation{%
  \institution{ETH Zürich}
  \country{Switzerland}
}

\pagestyle{fancy}
\fancyhf{} 
\renewcommand{\headrulewidth}{0pt}


\renewcommand{\headrulewidth}{0pt}

\maketitle

\enlargethispage{2\baselineskip}

\newcommand{\iscaheight}{0mm}
\ifdefined\eaopen
\renewcommand{\iscaheight}{12mm}
\fi

\section{Motivation}
\label{sec:motivation}

Recent advances in biotechnology and sensing technologies have enabled high-throughput, low-cost, and accurate biological data generation.
Modern sequencing platforms\nmg{~\cite{\citesbs,\citesmrt,\citenanopore}} and other omics technologies~\cite{dai2022advances} can generate massive amounts of biological data (genomics\nmg{~\cite{\citehwoptimization,\citealgoptimization}}, transcriptomics\nmg{~\cite{\citetranscriptomics}}, proteomics\nmg{~\cite{mallick2010proteomics,aslam2016proteomics,cho2007proteomics,patterson2003proteomics,graves2002molecular,pandey2000proteomics}}, and metabolomics\nmg{~\cite{alseekh2021mass,perez2019quantifying,liu2017metabolomics,zhang2012modern,johnson2012challenges}}) at rapidly decreasing costs. Similarly, multimodal medical imaging technologies\nmg{~\cite{marti2010multimodality,kasban2015comparative,shung2012principles,glover2011overview,kapoor2004introduction,townsend2004physical}}
produce high-resolution data that capture complex physiological and pathological processes. Wearable and implantable sensing devices\nmg{~\cite{pang2013recent,koydemir2018wearable,chan2012smart,lukowicz2004wearable,bonato2010wearable}} continuously monitor physiological signals such as heart activity, glucose levels, oxygen saturation in real time. Together, these advances have created an unprecedented volume of heterogeneous \kk{health- and life science-related} data.

This wealth of data creates unique opportunities for advancing healthcare and biomedical discovery, such as precision medicine~\cite{clark2019diagnosis,farnaes2018rapid,sweeney2021rapid,alkan2009personalized,flores2013p4,ginsburg2009genomic,chin2011cancer,Ashley2016}, where treatments and therapeutic strategies can be tailored to the genetic and physiological characteristics of individual patients. Large-scale genomic and clinical datasets also \nmg{advance 
personalized medicine~\cite{\citepersonalized}, tracking outbreaks of communicable diseases~\cite{\citeoutbreak}, cancer research~\cite{\citecancer}, bedside personalized care~\cite{chiang2019from}, agriculture~\cite{\citeagriculture}, ensuring food safety~\cite{e002244,TONG2021130},  scientific discovery~\cite{urbanek2018degradation,edgar2022petabase,paoli2022biosynthetic}, biodiversity conservation~\cite{Hogg2024,lewin2018earth}, evolutionary biology~\cite{\citeevolution}}.
%
Continuous physiological monitoring via wearable sensors further enables proactive and preventive healthcare by enabling early detection of anomalies~\cite{sopic2018real} and timely clinical intervention~\cite{guk2019evolution,tyler2020real}.

Despite these opportunities, efficiently analyzing large-scale biological data poses significant challenges for conventional computing systems. First, these systems often cannot keep up with the high-throughput rate at which data is generated. For example, modern sequencing platforms can generate data at rates that create substantial computational bottlenecks in downstream analysis, including \kk{sequence} alignment \kk{and} variant calling~\cite{zhang2021real,firtina2023rawhash,kovaka2020targeted, mutlu2023accelerating,Payne2021,Bao2021Squigglenet,ulrich2022readbouncer}. Similarly, images are generated at a pace that exceeds the throughput of  image reconstruction analytics and machine learning-based inference. High-throughput processing is crucial in clinical settings, where real-time data processing can significantly impact patient outcomes by improving both response times in time-critical scenarios and the decision-making processes for therapeutic schemes~\cite{kakria2015real}.

Second, \kk{health- and life science-related} applications suffer from data movement overheads\nmg{~\cite{wu2021sieve,shahroodi2022krakenonmem,shahroodi2022demeter,dashcam23micro,hanhan2022edam,zou2022biohd,cali2020genasm, huangfu2018radar, khatamifard2021genvom, gupta2019rapid, li2021pim, angizi2019aligns,zokaee2018aligner,Zhang_2023_alignerD,cali2022segram,kim2018grim,kaplan2020bioseal,mao2022genpip,alser2022molecules,mutlu2023accelerating,mansouri2022genstore,abakus23taco,megis,jun2016storage,kim2025nmp,soysal2025mars,zheng2025storage,grains,mansouri2026grains-extended,mutlu2025memory,mutlu2022modern,mutlu2019processing,mutlu2019enabling,senol2019nanopore}}. These workloads frequently involve irregular memory access patterns~\cite{yelick2020,langarita2022,kim2025nmp,robinson2021}, massive data transfer overheads~\cite{simon2026processing,ghose2019processing}, complex graph and statistical computations~\cite{li2022graph,yi2022graph}, and often rely on computationally intensive machine learning models~\cite{yuan2025ml}. As a result, data movement often dominates execution time and energy consumption. As datasets continue to grow in size and complexity, this bottleneck is further exacerbated~\cite{khan2020}. Growing adoption of deep learning models is also expected to amplify these challenges by requiring frequent movement of large model parameters and intermediate data across the computing stack~\cite{gholami2024}.

Third, systems must process sensitive patient data while satisfying strict privacy and regulatory requirements for health applications~\cite{rieke2020future,xu2021federated}. Hospitals, portable diagnostic devices and wearable platforms need to meet such requirements through secure data storage, trusted computation, federated learning, and privacy-preserving analytics~\cite{kaissis2020secure,sheller2020federated,zhang2025survey,kumar2025priv}. 

These challenges motivate the need for a closer collaboration between the \kk{health- and life science-related domain (e.g., clinical and precision medicine, omics research, genetics, computational biology, drug discovery, public health, and epidemiology)}, computer architecture, \kk{and computing systems design} communities. Advancing healthcare applications requires rethinking computing system design across the entire stack, including hardware accelerators, memory hierarchies, storage systems, data movement mechanisms, distributed infrastructures, and privacy-aware architectures. 
Prior work has already demonstrated the promise of this direction. A growing body of computer architecture research has proposed \kkthree{algorithmic optimizations (e.g.,\omcr{~\cite{\citealgoptimization,\citegraphalgoptimization,simon2026pim,altschul1990basic,breitwieser2022biodynamo,breitwieser2025design,breitwieser2023high,breitwieser2025teraagent}}) or hardware accelerators (e.g.,\omcr{~\cite{\citehwoptimization,\citegraphhwoptimization,simon2026pim,koliogeorgi2022gandafl,koliogeorgi2023profile,koliogeorgi2019dataflow,tsoutsouras2017exploration}}), and/or reducing data movement overheads\nmg{~\cite{mutlu2025memory,mutlu2022modern,mutlu2019processing,mutlu2019enabling,senol2021accelerating}} via \emph{near-data processing} (e.g., in main memory\nmgg{~\cite{singh2021fpga,wu2021sieve,shahroodi2022krakenonmem,shahroodi2022demeter,dashcam23micro,hanhan2022edam,zou2022biohd,cali2020genasm, huangfu2018radar, khatamifard2021genvom, gupta2019rapid, li2021pim, angizi2019aligns, zokaee2018aligner,Zhang_2023_alignerD,soysal2025mars,cali2022segram,kim2018grim,kaplan2020bioseal,mao2022genpip,shahroodi_swordfish_2023,alonso2024bimsa,diab2022high}} or storage\nmgg{~\cite{thesis,mansouri2022genstore,arxivGS,abakus23taco,megis,megisarxiv,jun2016storage,kim2025nmp,soysal2025mars,zheng2025storage,grains,grainsarxiv,mansouri2026sage,sagearxiv}})}
for workloads such as genome sequence analysis, metagenomic profiling, medical imaging, \kkthree{real-time physiological monitoring and biological simulation}, \kk{providing} substantial improvements in performance, energy efficiency, and scalability over conventional CPU- and GPU-based pipelines. 
Despite these promising advances, significant challenges remain unresolved \kk{and even bigger strides are necessary}.
Computer architecture research, \kk{done collaboratively with} \nmg{experts in healthcare and life sciences,} can play a key role in addressing these challenges and in enabling high-performance, energy-efficient, secure, and scalable \kk{computing systems for healthcare and life sciences}.
\section{The Arch4Health Initiative}
\label{sec:arch4health}

\textbf{Vision and Goals.}
We introduce the \emph{Architecture for Health} \emph{(Arch4Health)} initiative, which aims to (i)~identify key computational challenges in \kk{current and future} \kk{health- and life science-related} applications and (ii)~explore how computer architects \kk{and computing system designers} can advance healthcare by addressing these challenges. Since cross-disciplinary discussions are crucial for better identifying \kk{and solving} challenges in real-world \kk{health- and life science-related} applications, we aim to foster open discussions and cooperation between researchers with diverse backgrounds (i.e., from both computer architecture and health sciences communities, industry, and academia).

\noindent\textbf{Topics.}
Arch4Health invites contributions and collaborations on a broad range of topics at the intersection of computer architecture and \kk{health- and life science-related} applications. These topics include, but are not limited to, computational biology (e.g., genomics, metagenomics, \nmg{transcriptomics, single-cell analysis, spatial omics}, proteomics,  drug design and discovery, gene editing, and other areas in precision medicine \nmg{and public health}), neuroscience (e.g., brain-machine interfaces and prosthetics), wearable systems for health, medical robotics (e.g., surgery and haptics), mental health, medical imaging (e.g., brain scans, radiology, and single-cell analysis), \nmg{computational and digital pathology},
\nmg{artificial intelligence and foundation models for biology and health (e.g., protein
language models, cell foundation models, and large language models
for clinical applications)}, agent-based simulations, medical privacy, and bio-sensors. We particularly welcome contributions that identify computational challenges in these domains and propose new \kk{ideas (architectures and systems together with algorithms)} 
to improve performance, energy efficiency, cost-effectiveness, and privacy.

\noindent\textbf{Arch4Health Workshops.}
The primary instrument through which Arch4Health currently pursues its goals is a series of workshops co-located with various computer architecture and \kk{computing} systems conferences. \kk{These} workshops are designed to bring together researchers and practitioners from both the computer architecture \kk{and systems} community and the broader biomedical and healthcare communities in a single venue, and to encourage substantive technical exchange \kk{and visibility}. To this end, each workshop combines two complementary components. First, the workshops provide invited talks and keynotes that summarize \inum{i}~a series of research in computing system designs for healthcare applications and \inum{ii}~new \kk{ideas and directions} in data-intensive healthcare applications. Second, the workshops invite researchers to submit their ongoing work on these topics\kk{, with a focus on new ideas}. \kk{Third, all workshops are livestreamed and available on YouTube\nmgg{~\cite{onuryoutube,firsta4h,seconda4h}} to enable online, broad and unrestricted access to the entire world.}

The first two editions of Arch4Health were held in conjunction with the IEEE/ACM International Symposium on Microarchitecture (MICRO) 2025 in Seoul and the IEEE International Symposium on High-Performance Computer Architecture (HPCA) 2026 in Sydney, and together they engaged a broad community spanning computer architecture, bioinformatics, and biomedical research. The MICRO 2025 edition~\cite{arch4health-micro2025,firsta4h} featured ten invited talks covering new algorithms, software, and hardware designs for brain-computer interfaces, medical wearables, genomics, metagenomics, proteomics, and agent-based simulation in life sciences. The HPCA 2026 edition~\cite{arch4health-hpca2026,seconda4h} expanded the program to a full-day format with ten talks covering new algorithms, software, and hardware designs for genomics and metagenomics, electrical signals in genomics, AI and algorithms in biology, and indexing and querying petabyte-scale biological sequences. Across both editions, speakers came from a diverse set of academic and research institutions around the world.

\kk{The next edition of Arch4Health~\cite{arch4health-ics2026} \nmg{will be} held in conjunction with the ACM International Conference on Supercomputing (ICS) 2026 in Belfast, Northern Ireland, United Kingdom. This edition further expands the workshop series by fostering interdisciplinary collaboration among computer architecture, high-performance computing, and health and life science communities. The workshop program features invited talks and discussions on emerging algorithms for genome analysis, scalable architectures, hardware-software co-design for proteomics and genomics and real-time monitoring in clinical environments. We hope this edition will further solidify the importance of the initiative and lead to exciting synergies and research directions.}

\kktwo{Building on the success of the Arch4Health~\cite{arch4health-hpca2026,arch4health-micro2025} workshop series, we are expanding the scope toward the \kkthree{system software and system design} domain by organizing the Sys4Health workshop~\cite{sys4health-sosp2026}, co-located with the 32nd Symposium on Operating Systems Principles conference (SOSP 2026). Sys4Health aims to bring together the systems and health and life sciences communities to explore how systems and software infrastructure can enable scalable, secure, and efficient health and life science applications.}

\head{Future Outlook and a Call to Action}
The challenges and opportunities at the intersection of computer architecture and health- and life science-related domains are broad and pressing. Therefore, realizing the \kk{vision} and goals of Arch4Health requires diverse and cross-disciplinary discussions. We invite computer architects \kk{and computing system designers} to view health-related applications as a rich and important domain for architectural innovation, and we invite biomedical researchers and clinicians to engage with the architecture community in shaping the systems that will support the next generation of healthcare. 

Arch4Health is intended to grow as an open and inclusive initiative. We welcome contributions, ideas, and collaborations \kk{ across multiple communities}. \kk{Engagement and synergy between both \nmg{academia} and industry is essential to identify key challenges and pave the way toward impactful, practical solutions.} Through continued workshops, joint research efforts, and community-building activities, we hope Arch4Health will help grow \emph{architecture for health} as a key area of \kk{computer architecture}, whose advances translate into substantial \kk{impact} in advancing healthcare and life sciences. 


\begin{acks} 
\kk{This paper reflects the vision, objectives, and activities of the \textbf{Architecture for Health (Arch4Health)} workshop initiative.
We thank the SAFARI Research Group members for providing a stimulating intellectual and scientific environment. 
We acknowledge the generous gifts from our industrial partners, including Google, Huawei, Intel, and Microsoft. This work, along with our broader work in studying and accelerating health- and life-science-related applications \kkthree{(e.g.,~\cite{alser2020accelerating,cali2017nanopore,singh2021fpga,alser2022molecules,mutlu2023accelerating, mutlu2022modern, mutlu2025memory, soysal2025mars,firtina2023rawhash,firtina2023rawhash2,firtina_rawsamble_2024,lindegger2023rawalign,rawbench,megis,mansouri2026sage,grains,simon2026pim,diab_framework_2022,shahroodi2023swordfish,shahroodi_demeter_2022,cali2020genasm,cali2022segram,cavlak2024targetcall,kim_airlift_2024,kim2018grim,singh2024rubicon,firtina_blend_2023,firtina2020apollo,Lindegger2023scrooge,eudine2026genpairx,Pavon2024quetzal,alser2020sneakysnake,alser2017gatekeeper,bingol2021gatekeeper,thesis,alser2019shouji,kim_grim-filter_2018,alonso2024bimsa,firtina2024aphmm,kim2022fastremap,lindegger2022algorithmic,breitwieser2022biodynamo,breitwieser2023high,breitwieser2025design,breitwieser2025teraagent,firtina2025enabling,senol2021accelerating,koliogeorgi2023hardware})}, is supported in part \kktwo{by the European Union’s Horizon Program for research and innovation under Grant  No. 101047160 (project BioPIM), the Swiss National Science Foundation (SNSF) under Grant No. 213084}, the Semiconductor Research Corporation (SRC), the ETH Future Computing
Laboratory (EFCL), ACCESS – AI Chip Center for Emerging Smart Systems, and the Microsoft Swiss Joint Research Center. No AI or LLM help was used in creating this work.}
\end{acks} 




\bibliographystyle{ACM-Reference-Format}
\bibliography{refs}


\end{document}